\documentclass[epjCONF, onecolumn]{svjour}
\usepackage{graphicx}
\usepackage[varg]{txfonts} % Times fonts
\usepackage[latin1]{inputenc}
\usepackage{natbib}
\session-title{Assembling the Puzzle of the Milky Way}

\def\aj{AJ}%% Astronomical Journal
\def\apj{ApJ}%% Astrophysical Journal
\def\mnras{MNRAS}%% Monthly Notices of the RAS
\begin{document}
\title{Using frequency maps to constrain the distribution function of the Milky Way stellar halo}
\author{Monica Valluri\inst{1}\fnmsep\thanks{\email{mvalluri@umich.edu}}}
\institute{Department of Astronomy, University of Michigan, Ann Arbor 48109, USA.}
\abstract{Resolved surveys  of the Milky Way's stellar halo can obtain all 6 phase space coordinates of tens of thousands of individual stars, making it possible to compute their 3-dimensional orbits. Spectral analysis of large numbers of halo orbits can be used to construct frequency maps which are a compact, yet informative representation of their phase space distribution function (DF). Such maps can be used to infer the major types of orbit families that constitute the DF of stellar halo and their relative abundances. The structure of the frequency maps, especially the resonant orbits, reflects the formation history and shape of the dark matter potential and its orientation relative to the disk. The application of frequency analysis to cosmological hydrodynamic simulations of disk galaxies shows that the orbital families occupied by halo stars and dark matter particles are very similar, implying that stellar halo orbits can be used to constrain the DF of the dark matter halo, possibly impacting future direct dark matter detection experiments. An application of these methods to a sample of $\sim 16,000$  Milky Way halo and thick disk stars from the SDSS-SEGUE survey yields a frequency map with strong evidence for resonant trapping of halo stars by the Milky Way disk, in a manner predicted by controlled simulations in which the disk grows adiabatically. The application of frequency analysis methods to current and future phase space data for Milky Way halo stars will provide new insights into the formation history of the different components of the Galaxy and the DF of the halo.
} %end of abstract
\titlerunning{Frequency maps of the stellar halo}
\authorrunning{Valluri, Monica}
\maketitle
\section{Introduction}
\label{intro}

The three fundamental frequencies ($\Omega_1, \Omega_2, \Omega_3$) of oscillation of an orbit can be accurately extracted using a Fourier Transform method known as frequency analysis \citep{laskar_93, valluri_merritt_98}. These orbital frequencies can be used to obtain a complete picture of the properties of individual orbits as well as the entire phase space distribution function  (DF)~
 \citep{valluri_etal_11a}. Frequency analysis can be used to distinguish between regular and chaotic orbits,  to classify regular orbits into major orbit families, and to quantify the average shape of  an orbit and relate its shape to the shape of the halo \citep{valluri_etal_10}. Resonant orbits are regular orbits that have fewer than 3 linearly independent fundamental frequencies which are related via integer
linear equation such as: $l\Omega_1+m\Omega_2+n\Omega_3 = 0,$ where
($l,m,n$) are small integers.  A frequency map constructed from pairs of frequencies (e.g. $\Omega_1/\Omega_3$ vs. $\Omega_2/\Omega_3$) can be used to easily
identify the most important resonant orbit families, since such orbits
populate straight lines on such a map. The strength (or importance) of
a resonance can be assessed from the number of orbits associated with it.  In this contribution will briefly summarize recent applications of the frequency analysis method to both controlled and cosmological simulations of disk galaxies and to real stellar orbits derived from phase space coordinates obtained with the SDSS SEGUE survey.%, to  illustrate the power of this method and discuss how it can aid in the construction of the DF of the Milky Way.

\section{Effect of  dark matter halo shape and orientation on orbital properties}
\label{sec:frequency_maps}

\begin{figure}
\begin{center}
% For example, with the option graphics use
%\resizebox{0.75\columnwidth}{!}{%
\includegraphics[trim=0.pt 0.pt 0.pt 0pt,width=0.31\textwidth]{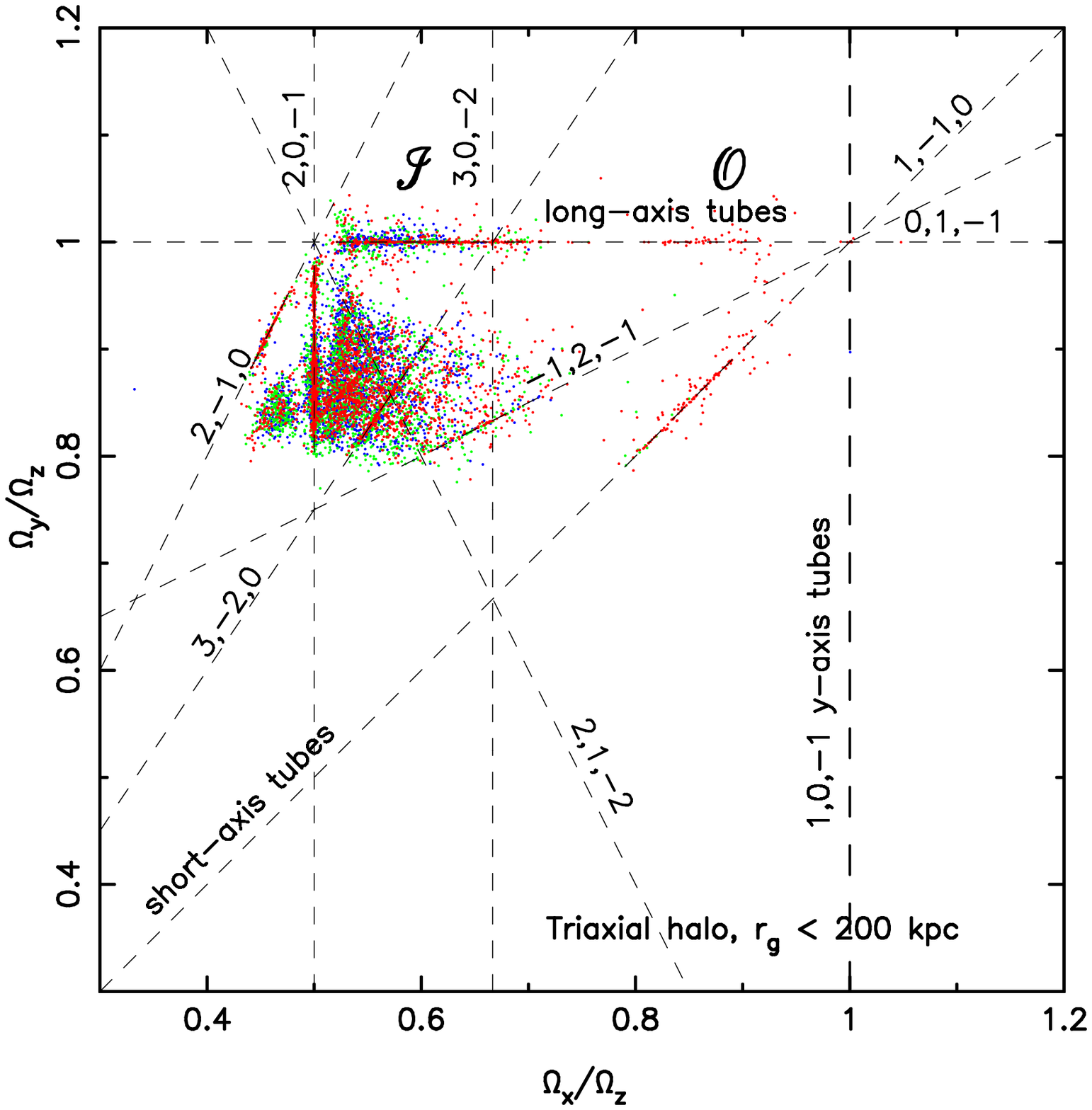}
\includegraphics[trim=0.pt 0.pt 0.pt 0pt,width=0.31\textwidth]{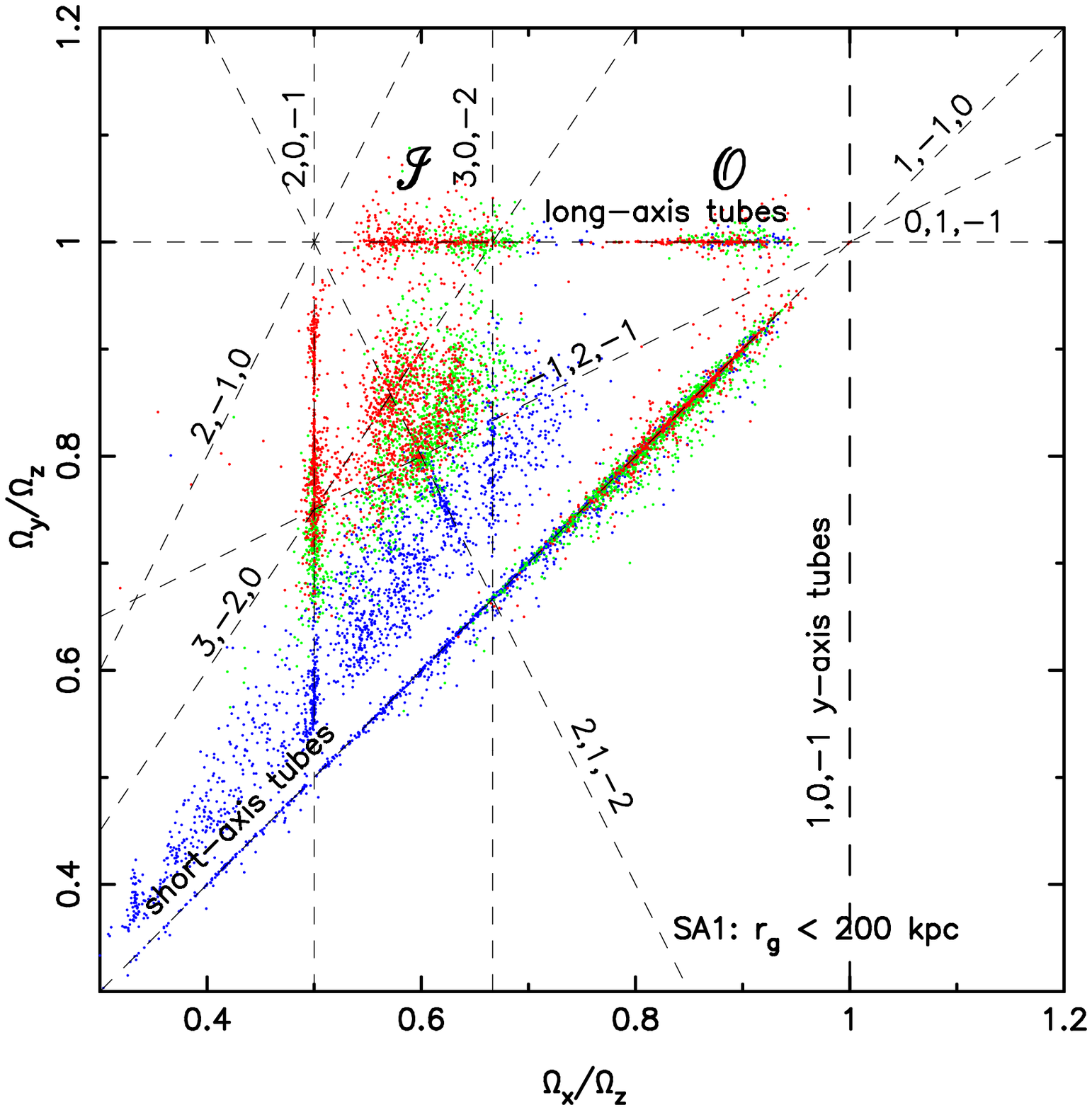} 
\includegraphics[trim=15.pt 0.pt 0.pt 0pt,clip,width=0.32\textwidth]{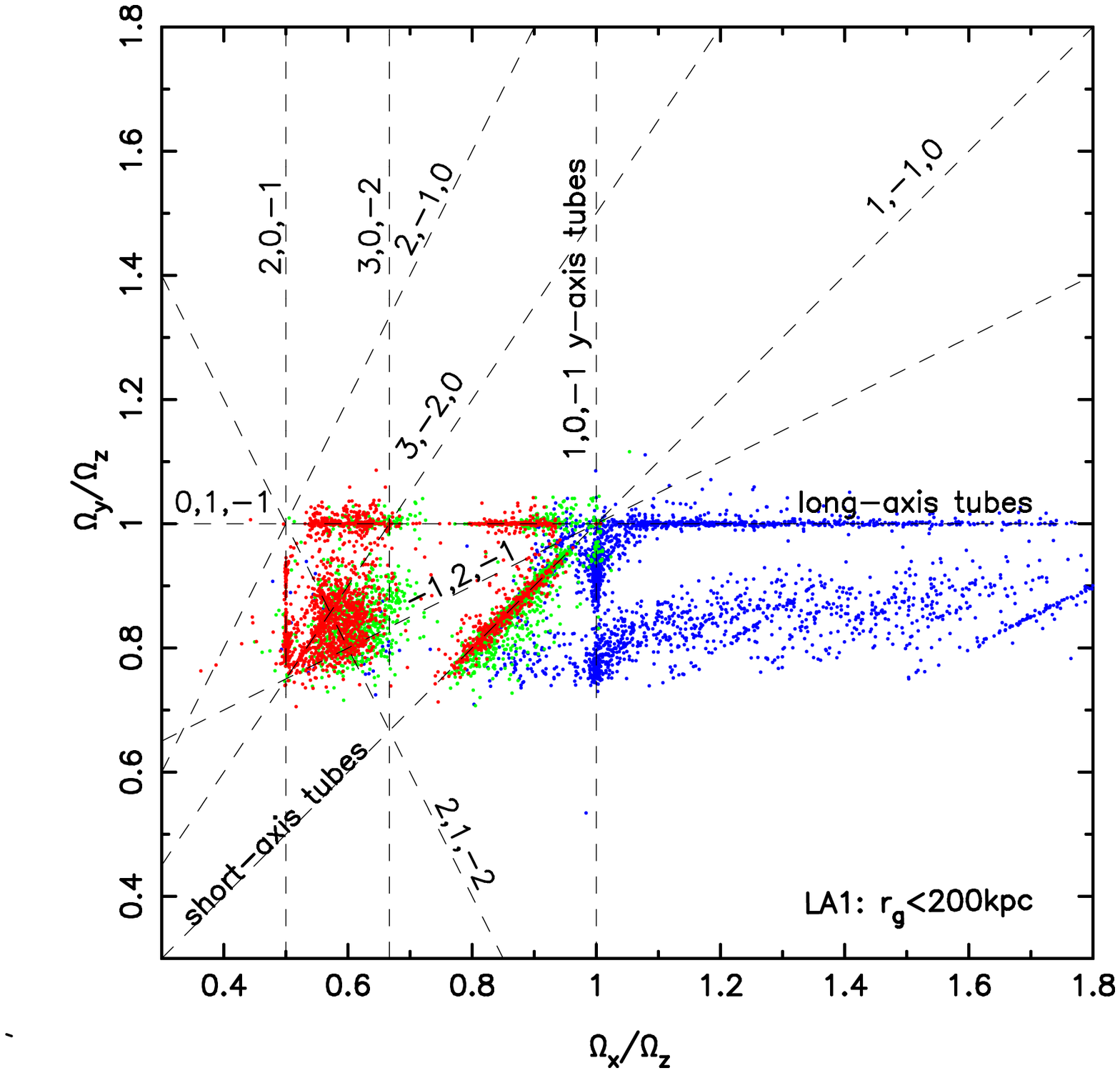} 
\end{center}
\vspace{-0.5cm}
\caption{Frequency maps of $\sim 10^4$ halo orbits with  radius $ <200~$kpc, Left:  in a  triaxial halo model; Middle: disk perpendicular to halo short-axis (model SA1); Right: disk perpendicular to halo long-axis (model LA1). Dashed lines mark resonances characterized by integers $(l,m,n)$;  inner and outer long-axis tubes marked by script I and O along $\Omega_y/\Omega_z \sim 1$. 
 % {\it Short-axis tubes} are along the diagonal line with $\Omega_x/\Omega_y \sim 1$. 
Color coding is by energy (blue: tightly bound; green: intermediate; red: weakly bound).}
\label{fig:fig1}       % Give a unique label
\end{figure}

Controlled simulations show that the orbital content of a halo is modified adiabatically by the growth of a baryonic component \citep{valluri_etal_10}.  When ratios of orbital fundamental frequencies $\Omega_x, \Omega_y, \Omega_z$ (in Cartesian coordinates) of a large number ($1-2\times 10^4$) halo orbits from a {\it self-consistent DF} are plotted in a frequency map (Fig.~\ref{fig:fig1}), it represents the orbital content of the DF. %The particles are color coded by binding energy: the 1/3rd most weakly bound are red, the 1/3 most tightly bound are blue and the rest are colored green.
 For a strongly triaxial halo (created via multiple mergers of spherical NFW halos) the frequency map (Fig.~\ref{fig:fig1}[left]) shows numerous points clustered along the horizontal line at $\Omega_y/\Omega_z=1$ corresponding to long-axis tubes (L-tubes).  The clump of points below this line corresponds to non-resonant box orbits and the vertical line at $\Omega_x/\Omega_z=0.5$ corresponds to resonant  boxlet (banana) orbits. Other resonant box orbit families are also populated. \citet{valluri_etal_11a} show that when a disk galaxy grows inside such a  triaxial halo, in addition to changing the halo's shape, it causes significant resonant trapping of halo orbits. The specific orbit families that are trapped depend strongly on the {\it relative orientation} of the disk and the halo. In Figure~\ref{fig:fig1}[middle] the disk plane is perpendicular to the short axis of the original triaxial halo (on left) and therefore traps halo orbits into the short-axis tube  (S-tubes) family (seen as the enhanced clustering along diagonal line with $\Omega_x/\Omega_y \sim 1$). This DF (model SA1) is dominated by box orbits (48\%) and S-tubes (32\%), with L-tubes (12\%) and chaotic orbits (8\%) playing only a small role.  In contrast when the disk grows perpendicular to the long-axis of the halo (Fig.~\ref{fig:fig1}[right]) it traps orbits into the L-tube family (seen as enhanced clustering along the horizontal line with $\Omega_y/\Omega_z=1$). Thus the relative importance of orbits associated with L-tubes versus S-tubes is reflected by the increased strength of the orbit family that shares the symmetry of the disk. Thus a frequency map reflects the {\it abundance of these orbits in the final DF  and the relative orientation} of the disk and halo. %We also showed  \cite{valluri_etal_11a} that these results hold even when the initial positions of stars are selected to lie within 10~kpc of the Sun.

\section{Orbits of stars and dark matter in cosmological simulations}
\label{sec:MUGS}
 
\begin{figure}
\begin{center}
% Use the relevant command for your figure-insertion program
% to insert the figure file.
% For example, with the option graphics use
%\resizebox{0.75\columnwidth}{!}{%
\centering \includegraphics[trim=0.pt 0.pt 0.pt 0pt, angle=-90,width=0.33\textwidth]{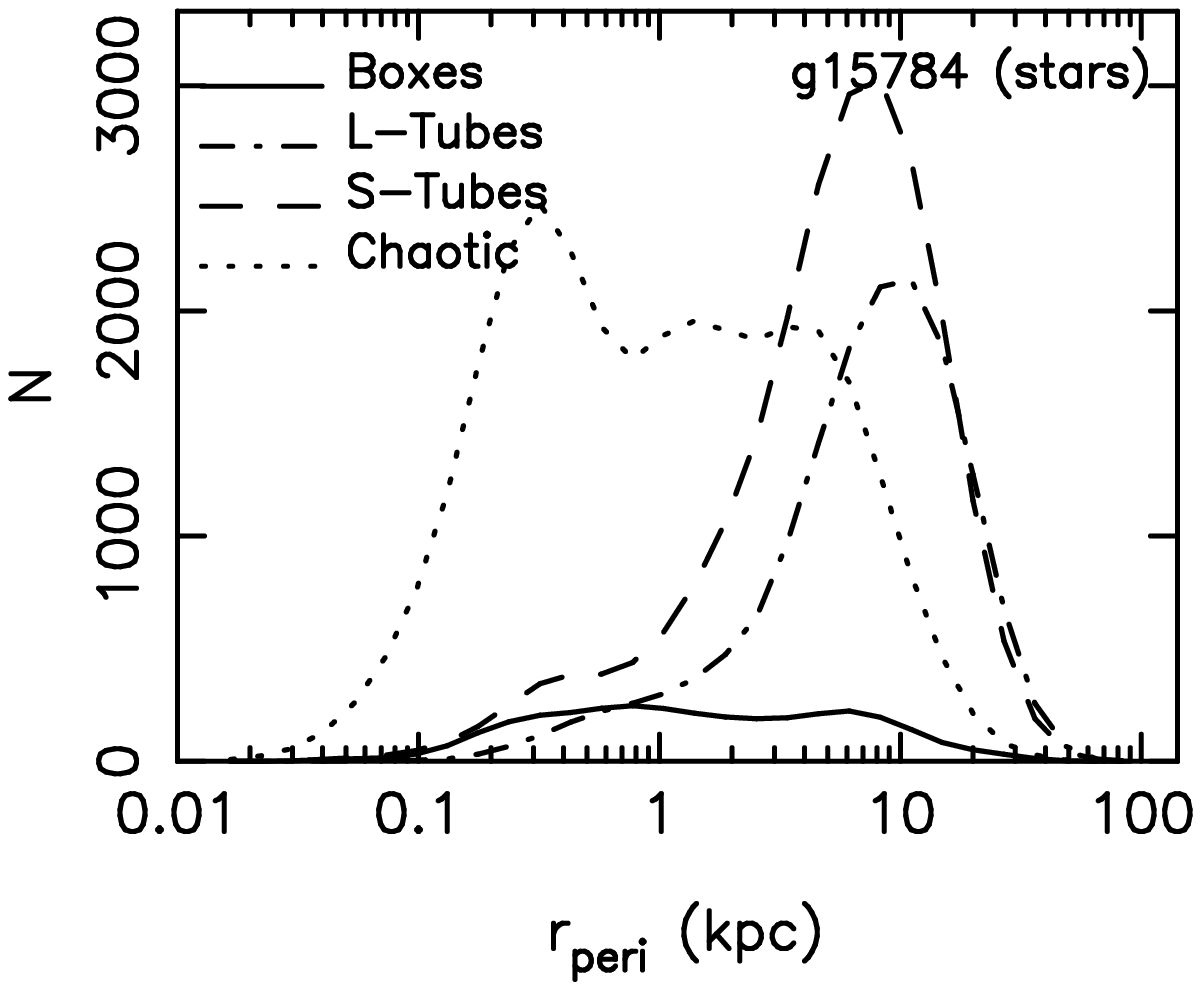}
\centering \includegraphics[trim=0.pt 0.pt 0.pt 0pt,angle=-90,width=0.33\textwidth]{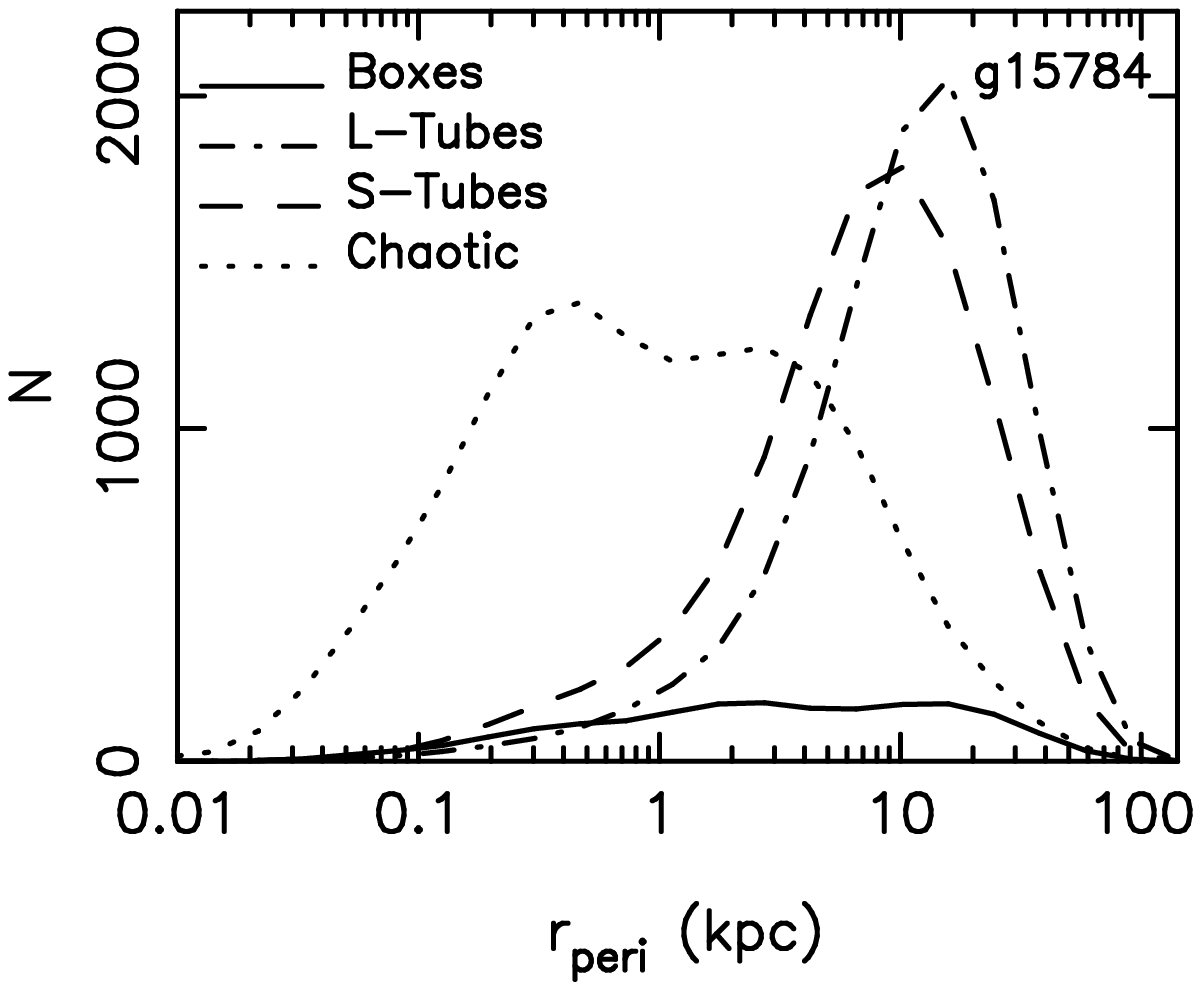}
\end{center}
\vspace{-0.5cm}
\caption{Distribution of orbit types as a function of orbital
  pericenter distance $r_{peri}$ for orbits in a cosmological simulation of a disk galaxy (g15784): halo stars (left) and halo dark matter particles (right). }
\label{fig:fig2}       % Give a unique label
\end{figure}

 \citet{valluri_etal_11b} analyze a simulation of a disk galaxy in a fully cosmological context \citep[part of the McMaster Unbiased Galaxy Simulations sample, hereafter MUGS,][]{stinson_etal_10}. The simulations include low-temperature metal cooling, heating from UV background radiation,
star formation, stellar energy and metal feedback, and metal
diffusion, and produce fairly realistic disk galaxies. 
We use frequency analysis to study the orbit populations of both star particles and dark matter particles in the MUGS disk galaxy g15784. For this galaxy the dark matter halo is nearly oblate with  flattening parameter $q \sim 0.85$ and the stellar halo is also oblate, but flatter with $q \sim 0.7$ so one would expect S-tubes to dominate the DF. Figure~\ref{fig:fig2} shows histograms of orbits of different types as a function of the orbital pericenter radius $r_{peri}$. The box orbits are unimportant ($<5$\%) for both stars and dark matter particles, and the inner part of the halo is dominated by chaotic orbits (42\% of stars, and 38\% of dark matter). At large radii both L-tubes (21\% for stars and 28\% for dark matter) and S-tubes (32\% for stars and 29\% for dark matter) are equally important, contrary to what might be expected  simply from the shape of the halo.  The similarity between the orbit populations of stars and DM particles is striking, but there are subtle differences, with slightly more stars on S-tubes (with symmetry axis of the disk) and slightly larger numbers of DM particles on L-tubes, indicating that they experienced low-angular momentum radial infall \citep{valluri_etal_10}.  %Analysis of subsets of the orbits, sliced by their maximum excursion above the disk or by stellar metallicity show no 

\section{Evidence for resonant trapping of halo stars in SEGUE data}
\label{sec:SEGUE}

\citet{oshea_etal_11} have recently analyzed the orbital properties of a subset of stars
from the Sloan Digital Sky Survey \citep[SEGUE][]{SEGUE} whose three dimensional velocities
are well-quantified.  This set of stars was recently used to show that the stellar halo may consist of two distinct but overlapping structural components with slightly different rotation and  metal content \citep{carollo_etal_10}. The sample consists of stars with $r-r_\odot < 4$kpc for which all 6 phase-space coordinates are available and includes thin and thick disk stars but is dominated by halo stars \citep{carollo_etal_10}. We used the phase space coordinates to integrate the orbits of these stars in a Milky Way potential modeled by an oblate axisymmetric NFW dark matter halo ($q=0.95$), a Hernquist bulge, and Miyamoto-Nagai disk with parameters that match the rotation curve of the Milky Way.  Since the potential is flattened and axisymmetric, frequency analysis in cylindrical coordinates yields  radial, azimuthal and vertical oscillations frequencies  $\Omega_R$, $\Omega_\phi$ and $\Omega_z$. %This is analogous to decomposing motions in stellar disks into radial (epicyclic), circular and vertical oscillation \cite{BT}, although the oscillations are not small for halo stars. 
The sign of $\Omega_\phi $ is either positive or negative, depending on whether the particle circulates clockwise or anti-clockwise about the Galactic center.  

\begin{figure}
\begin{center}
% Use the relevant command for your figure-insertion program
% to insert the figure file.
% For example, with the option graphics use
%\resizebox{0.75\columnwidth}{!}{%
%  \includegraphics{fig1.eps} }
\includegraphics[trim=0.pt 0.pt 0.pt 0pt,width=0.31\textwidth]{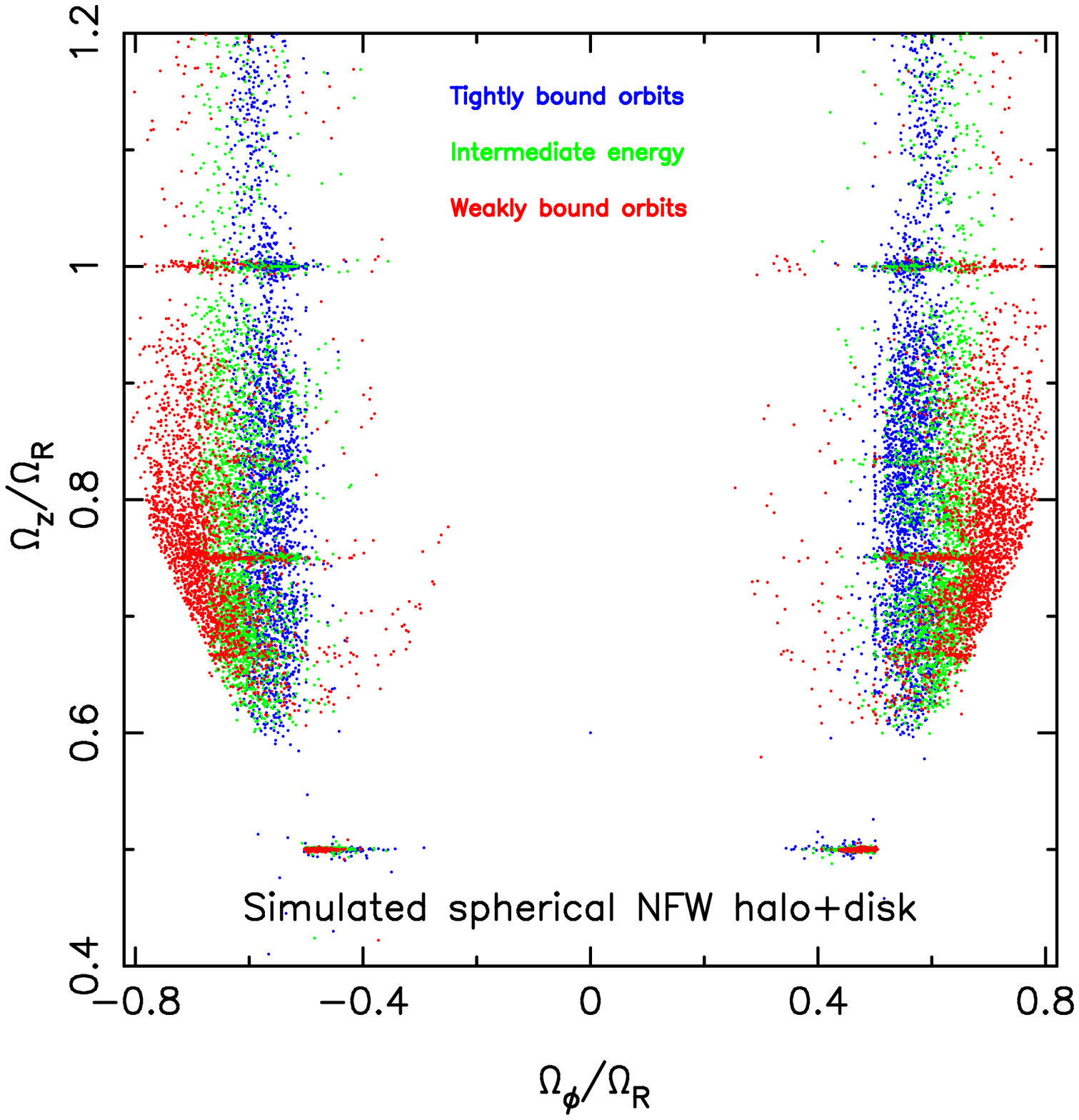}
\includegraphics[trim=0.pt 0.pt 0.pt 0pt,width=0.32\textwidth]{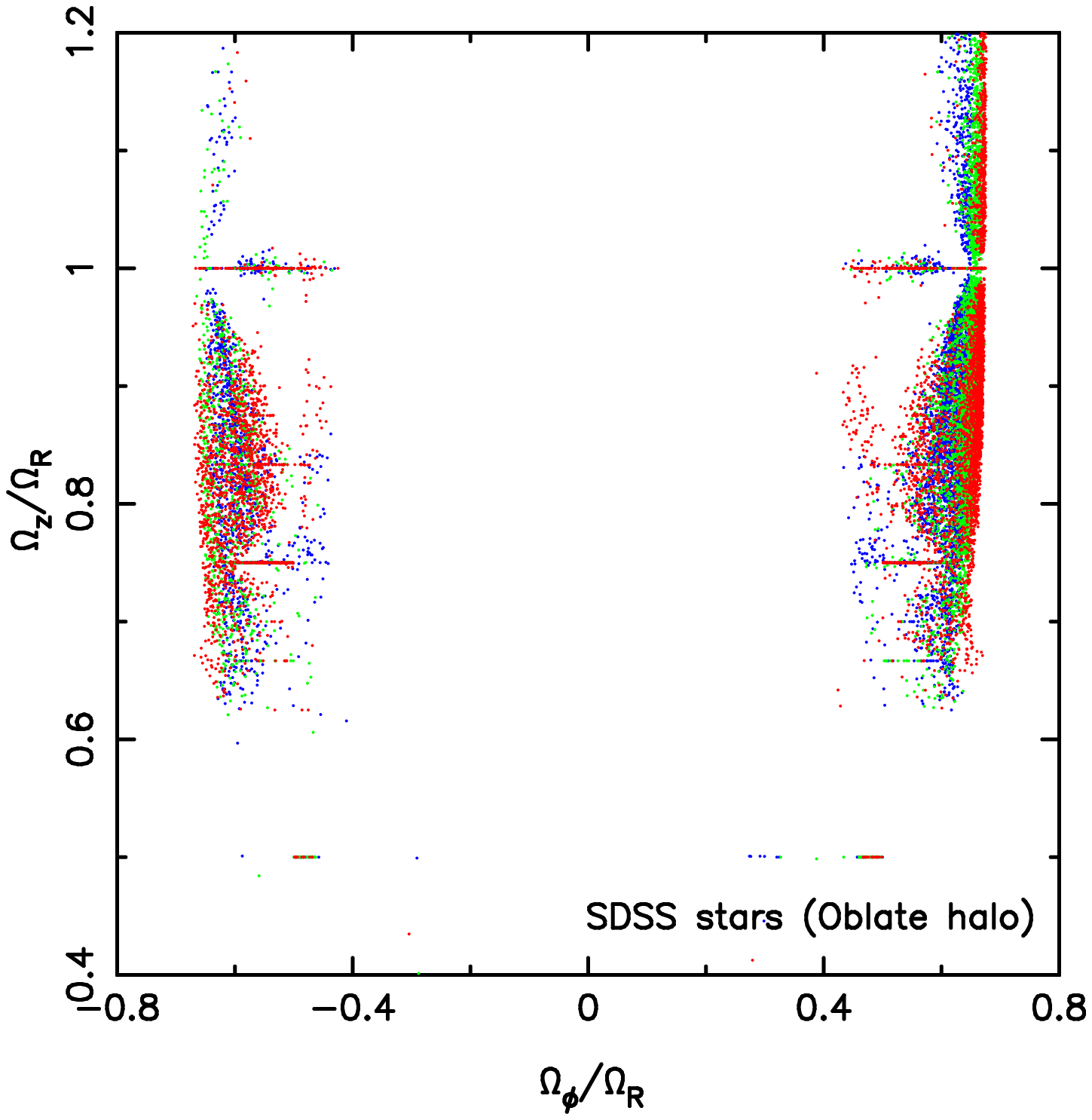}
\includegraphics[trim=0.pt 0.pt 0.pt 0pt,width=0.32\textwidth]{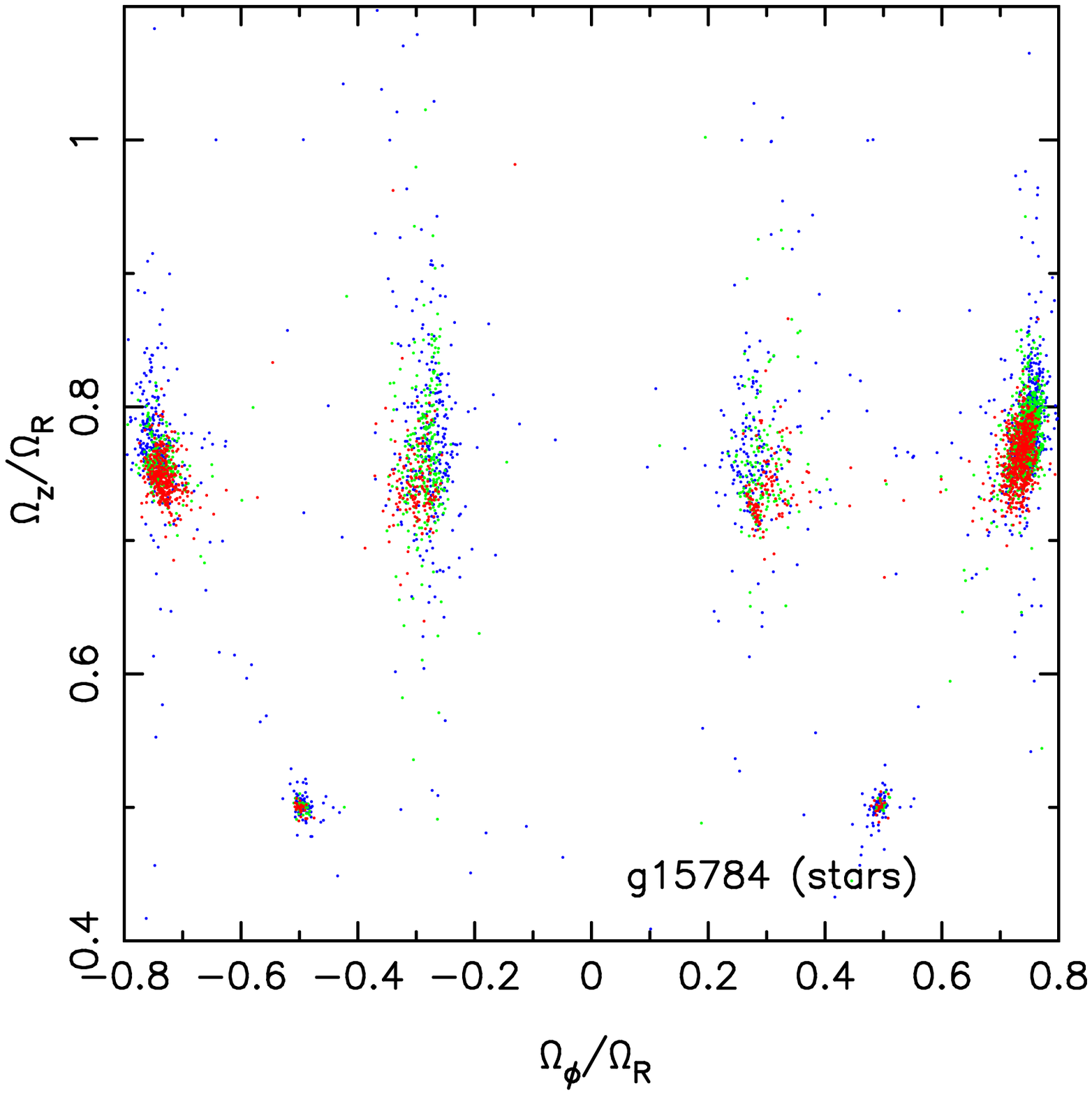}
\end{center}
\vspace{-0.5cm}
\caption{ Left: Frequency maps of $10^4$ halo orbits in the simulation where a stellar disk grows in a spherical NFW halo. The bisymmetry about $\Omega_\phi/\Omega_R=0$ results because the  simulated halo has no net rotation. The map shows several resonances which appear as horizontal
lines: e.g. $\Omega_z/\Omega_R=1/2, 2/3, 3/4, 5/6, 1, 3/2$. Middle: Frequency map of orbits of stars from the SDSS-SEGUE sample evolved in an axisymmetric potential: resonances are seen at $\Omega_z/\Omega_R =1/2, 3/4, 5/6, 1$. Right: Frequency map of $10^4$ halo stars from MUGS galaxy g15784. Only one resonance is seen at $\Omega_z/\Omega_R=1/2$}
\label{fig:fig3}       % Give a unique label
\end{figure}

To provide a basis for understanding the frequency map in cylindrical coordinates Fig.~\ref{fig:fig3}[left]  shows  the orbits of 20,000 randomly selected halo particles in a controlled simulation in which a disk galaxy was grown inside a spherical NFW halo with no net rotation \citep{valluri_etal_11a}. The growth of the disk inside this halo, made it oblate with flattening $q \sim 0.9$ similar to that assumed for our Milky Way model.  This controlled adiabatic simulation shows  symmetry about $\Omega_\phi/\Omega_R=0$ because the orbits were selected at random from a DF with no net rotation. 
The frequency map of 16,000 calibration stars from the SDSS-Segue survey is shown in Fig.~\ref{fig:fig3}[middle], while  Fig.~\ref{fig:fig3}[right] shows  a map of halo stars from the MUGS disk galaxy g15784 (see \S~3).  In the left and middle maps points lie in two distinct lobes, corresponding to  orbits with positive and negative angular momentum. In addition they cluster along several thin horizontal lines corresponding to orbital resonances $\Omega_z/\Omega_R$ = 0.5 (1:2), 0.66 (2:3), 0.75 (3:4), 0.83 (5:6), 1 (1:1 and 2:2). The SEGUE data [middle panel] shows a larger fraction of stars in the right lobe, corotating with the disk (confirming the finding of \cite{carollo_etal_10}). Although there are some differences, the similarities in the strengths of the resonances and the specific resonances ratios in the left and middle panels are striking. Figure~\ref{fig:fig3}[left] shows that the phase space DF of orbits of stars selected from simulated disk g15784 is strikingly different from the other two DFs. This galaxy experienced its last major merger $z=2$ but is perturbed even at $z=0$ by scattering of stars by dwarf companions and  dark matter subhalos destroys resonances \citep{valluri_etal_11b}. Although the shapes of the potentials in all three systems are very similar, the differences in the frequency maps reflect differences in the {\it distribution functions}, demonstrating the power of frequency mapping to uncover the halo's phase space structure from samples of a few $\times 10^4$ halo orbits. 
 
\section{Conclusions}
\label{sec:conclusions}
 Previous applications of frequency mapping have been restricted to orbits that uniformly sampled initial condition space to study the structure of the underlying potential. Recent applications to large samples of orbits drawn from a {\it self-consistent distribution function} show that this is a compact way to represent the full phase space DF \citep{valluri_etal_11a} which separates orbits into their major families, and highlights those that have been resonantly trapped by the formation process.  The strong similarities between the distribution functions of a controlled adiabatic simulation of a disk galaxy in a spherical halo and the real orbits derived from SEGUE, suggest that resonant trapping of halo stars by the  Milky Way's disk could have occurred in the past few Giga years, providing evidence for a relatively quiescent recent history. A study of dark matter and stellar orbits in a cosmological hydrodynamical simulation of a disk galaxy shows relatively insignificant differences between these two components suggesting that with large samples of halo orbits, we may be able to gain insights into DF of dark matter particles themselves, significantly impacting future direct dark matter detection searches.

\section*{Acknowledgments}
MV is supported by NSF grant AST-0908346 and thanks her collaborators particularly J. Bailin, T. Beers, V.P. Debattista,  B. O'Shea, T.Quinn and G. Stinson  for permission to summarize the results of various   applications of spectral analysis methods, some prior to publication.

%\begin{thebibliography}{}
% and use \bibitem to create references.
%\bibitem{RefJ}
% Format for Journal Reference
%Author, Journal \textbf{Volume}, (year) page numbers
% Format for books
%\bibitem{RefB}
%Author, \textit{Book title} (Publisher, place year) page numbers
% etc
%\end{thebibliography}
%\bibliography{Master}

\end{document}